\documentclass[useAMS,usenatbib]{mn2e}
\usepackage{graphicx,psfig,cite,epsf}  
\usepackage{times}
\usepackage{subfigure}
\usepackage{textcomp}
\usepackage{graphicx,times,multirow,amsmath}
\usepackage{natbib}

\def\LaTeX{L\kern-.36em\raise.3ex\hbox{a}\kern-.15em
    T\kern-.1667em\lower.7ex\hbox{E}\kern-.125emX}

\def\cxo {\emph{Chandra}}
\def\swift {\emph{Swift}}

\def\src {NGC\,55~ULX1}

\def\nh {$N_{\rm H}$}

\def\apjl{ApJ}
\def\apjl{ApJ}                % Astrophysical Journal, Letters
\def\apjs{ApJS}               % Astrophysical Journal, Supplement
\def\ao{Appl.~Opt.}           % Applied Optics
\def\aap{A\&A}                % Astronomy and Astrophysics
\title[Spectral variability of the Ultraluminous source NGC 55 ULX1]{Spectral variability in Swift and Chandra observations of the Ultraluminous source NGC 55 ULX1}

\author[F. Pintore, P. Esposito, L. Zampieri, S. Motta, A. Wolter]{Fabio Pintore$^1$, Paolo Esposito$^{2,3}$, Luca Zampieri$^4$, Sara Motta$^5$, Anna Wolter$^6$\\
$^1$ Dipartimento di Fisica, Universit\`a di Cagliari, I-09042 Monserrato, Italy\\
$^2$ Istituto di Astrofisica Spaziale e Fisica Cosmica - Milano, INAF, via E. Bassini 15, I-20133 Milano, Italy\\
$^3$ Harvard--Smithsonian Center for Astrophysics, 60 Garden Street,
Cambridge, MA 02138, USA\\
$^4$ INAF-Osservatorio Astronomico di Padova, Vicolo dell'Osservatorio 5, I-35122 Padova, Italy \\ 
$^5$ ESAC, European Space Astronomy Centre, Villanueva de la Ca$\tilde{\text{n}}$ada, E-28692 Madrid, Spain\\
$^6$ INAF, Osservatorio Astronomico di Brera, via Brera 28, 20121 Milano, Italy\\}

\begin{document}

\date{Accepted ... Received ...; in original form ...}
\pagerange{\pageref{firstpage}--\pageref{lastpage}} \pubyear{2010}
\maketitle

\label{firstpage}

\begin{abstract}
NGC 55 ULX1 is a bright Ultraluminous X-ray source located 1.78 Mpc away. We analysed a sample of 20 \textit{Swift} observations, taken between 2013 April and August, and two \textit{Chandra} observations taken in 2001 September and 2004 June. We found only marginal hints of a limited number of dips in the light curve, previously reported to occur in this source, although the uncertainties due to the low counting statistics of the data are large. The \textit{Chandra} and \textit{Swift} spectra showed clearly spectral variability which resembles those observed in other ULXs. We can account for this spectral variability in terms of changes in both the normalization and intrinsic column density of a two-components model consisting of a blackbody (for the soft component) and a multicolour accretion disc (for the hard component). We discuss the possibility that strong outflows ejected by the disc are in part responsible for such spectral changes.
\end{abstract}


\section{Introduction}
\label{sect1}

The population of Ultraluminous X-ray sources (ULXs) has grown significantly after the launch of the \emph{XMM-Newton} and \emph{Chandra} satellites. These observatories made it possible to carry out the detailed spectroscopic study of intrinsically powerful but faint point-like X-ray sources in nearby galaxies ($<100$ Mpc). They were dubbed Ultraluminous X-ray sources because they have a very high X-ray luminosity, above $10^{39}$ erg s$^{-1}$ \citep{fabbiano89}, and exceed (although not necessarily all the time) the Eddington limit for spherical accretion of pure hydrogen onto a 10 M$_{\odot}$ black hole (BH). Their position is not coincident with the nucleus of their host galaxy, ruling out a low luminosity Active Galactic Nucleus (AGN). While a significant fraction of ULXs can be associated to background AGNs or young supernovae interacting with the circumstellar matter, the majority of them are now thought to be a peculiar class of accreting sources, either BHs of stellar origin (e.g. \citealt{king01}) accreting above Eddington or Intermediate Mass BHs (IMBHs, e.g. \citealt{colbert99}) accreting sub-Eddington. Because of the special conditions/environment required to form IMBHs \citep{madau01,portegies04,vandermarel04}, the latter scenario is considered a viable one for HLX-1 (\citealt{farrell09,servillat11}), that may however be interpreted also as the nucleus of a stripped dwarf galaxy (\citealt{soria11b,webb10, mapelli12, mapelli13}), and possibly for a handful of other sources (e.g. \citealt{sutton12,pasham14}). 
On the other hand, for the great majority of ULXs an interpretation in terms of a stellar mass BH or a BH of stellar origin seems more likely (see e.g. \citealt{zampieri09, fengsoria11}). Recent dynamical measurements or constraints to the BH mass have been derived and are in agreement with this conclusion \citep{liu13, motch14}. In addition, the observed X-ray spectral and variability properties are consistent in most cases with an interpretation in terms of super-Eddington accretion and possibly beamed emission (e.g. \citealt{gladstone09, sutton13}), as expected for light BHs, with the actual BH mass (5-80 M$_{\odot}$) depending somewhat on the metallicity of the environment (e.g. \citealt{zampieri09, mapelli09, belczynski10}). Finally, it was very recently shown that some transient ULXs may even host Neutron Stars \citep{bachetti14}. 

One of the main diagnostics to study ULXs is the X-ray spectrum. High quality ULX spectra are usually characterised by a soft component, well fit by a standard disc spectrum, and an additional high energy component with a turn-over at 3-5 keV, phenomenologically described in terms of  Comptonization in an optically thick medium (\citealt{stobbart06,gladstone09,pintore12}). Although this two-components model is only a phenomenological description, the peculiar properties of the ULX spectra has led to the conclusion that these sources are in a markedly different accretion regime from those observed in Galactic BH X-ray binary systems (XRBs) accreting at sub-Eddington rates (\citealt{gladstone09,kajava09}). The combination of spectral shape, spectral evolution and variability suggest that the {mass accretion rate in the disc is} at or above the Eddington limit \citep[e.g.][]{middleton11,sutton13,bachetti13,walton13,pintore14}. Under such conditions, strong, radiatively driven winds can originate from the inner regions of the disc and propagate outwards above its plane (e.g. \citealt{poutanen07,ohsuga07,ohsuga09}). These turbulent and optically thick winds may originate the soft spectral component observed at low energies and possibly also the extrinsic variability detected at higher energies \citep[e.g.][]{middleton11b}, while an advection-dominated disc, the emission of which may be comptonized, produces the high energy part of the spectrum \citep[e.g.][]{middleton11,middleton12,sutton13,walton13,pintore14}. The structure and emission properties of the inner disc may be affected also by BH spin and radiative transfer effects taking place in the disc atmosphere \citep[e.g][]{suleimanov02,kawaguchi03}. This picture has been confirmed by observations above 10 keV recently performed by \textit{NuSTAR} (\citealt{bachetti13,rana14,walton13,walton14b}).
Clearly, the onset of strong outflows would play an important role in modifying the structure of the accretion disc as they are expected to effectively remove energy and angular momentum from the system.

In this work, we present the results of a \textit{Swift} monitoring campaign of the source NGC 55 ULX1, located in a Magellanic barred spiral galaxy at a distance of 1.78 Mpc \citep{karachentsev03} and with X-ray luminosity in the range $8 \times 10^{38} - 2 \times 10^{39}$ erg s$^{-1}$. This low-luminosity ULX is peculiar because in two \emph{XMM-Newton} observations it showed energy-dependent dips, not commonly observed in ULXs (see also \citealt{grise13} and \citealt{lin13} for possible dips in NGC 5408 X-1 and 2XMM J125048.6+410743, respectively). The dips lasted for a few hundreds of seconds \citep{stobbart04} and appeared to be correlated with time variability at high energy. A later \emph{XMM-Newton} observation caught the source in a low flux state where dips were not present and the spectral properties were different {(Pintore et al in prep.)}. The dips and the associated variability at high energy may be related to turbulences or blobs of optically thick matter in the wind that occasionally encounter our line of sight and mask the emitting inner regions (Middleton, Heil, Pintore, Walton and Roberts, submitted). The study of this source appears then crucial to assess both the spectral variability and the onset of powerful outflows in ULXs. This fact, along with the goal of investigating the source long-term flux variability, motivated our request for a \textit{Swift} monitoring campaign.
Besides the \textit{Swift} observations, we analyzed and present here also two previously unpublished \textit{Chandra} observations.

The paper is structured as follows: in Section~\ref{sect2} we summarize the data selection and reduction procedures, in Section~\ref{spectral_results} we present the results of our timing and spectral analysis of NGC 55 ULX1, and in Section~\ref{discussion} we discuss them.

 \begin{table}
\centering \caption{Summary of the \cxo\ and \textit{SWIFT} observations of NGC 55 ULX1.} \label{logs}
\scalebox{0.84}{\begin{minipage}{22cm}
\begin{tabular}{@{}lcccc}
\hline
\\
\textit{Chandra} Obs.\,ID  & Start time & Stop time & Exp. & Net counts\\
 & \multicolumn{2}{c}{(Terrestrial Time)} & (ks) &\\
\hline
2255  (Chip I0) & 2001-09-11~06:25:05 & 2001-09-11~23:30:20 & 59.4  &9509\\
4744 (Chip I1) & 2004-06-29~01:48:01 & 2004-06-29~05:17:04 & 9.6  &3719\\
\hline
\hline
\emph{Swift} Obs.ID & \multicolumn{3}{c}{}\\
\\
00032619001 & 2013-04-10 03:24:17 & 2013-04-10 09:50:56 & 4.8  & 301 $\pm 17$\\
00032619002 & 2013-04-17 11:22:00 & 2013-04-17 16:30:56 & 5.2  & 299 $\pm 18$\\
00032619003 & 2013-04-24 13:35:09 & 2013-04-24 23:19:54 & 4.8 & 178 $\pm 14$\\
00032619004 & 2013-05-01 00:48:38 & 2013-05-01 05:54:54 & 5.0  & 191 $\pm 14$\\
00032619005 & 2013-05-08 05:45:04 & 2013-05-08 23:54:57 & 4.9  & 258 $\pm 16$\\
00032619006 & 2013-05-15 18:48:28 & 2013-05-15 22:27:55 & 4.1  & 167 $\pm 13$\\
00032619007 & 2013-05-22 15:47:03 & 2013-05-22 22:49:55 & 5.6  & 182 $\pm 14$\\
00032619008 & 2013-05-29 19:32:23 & 2013-05-29 19:39:54 & 4.4  & 160 $\pm 10$\\
00032619009 & 2013-06-02 11:33:13 & 2013-06-02 15:04:55 & 4.5  & 360 $\pm 19$\\
00032619010 & 2013-06-05 00:18:17 & 2013-06-05 05:26:57 & 4.6  & 483 $\pm 22$\\
00032619011 & 2013-06-12 18:28:52 & 2013-06-12 23:25:56 & 3.4  & 168 $\pm 13$\\
00032619012 & 2013-06-19 18:28:08 & 2013-06-19 23:35:55 & 5.4  & 200 $\pm 14$\\
00032619013 & 2013-06-27 20:17:08 & 2013-06-27 23:54:56 & 4.9  & 170 $\pm 13$\\
00032619014 & 2013-07-03 00:03:17 & 2013-07-03 22:18:55 & 4.9  & 165 $\pm 13$\\
00032619015 & 2013-07-10 09:26:47 & 2013-07-10 14:40:55 & 5.2  & 327 $\pm 18$\\
00032619016 & 2013-07-17 11:14:55 & 2013-07-17 21:16:54 & 4.7  & 299 $\pm 17$\\
00032619017 & 2013-07-24 11:46:07 & 2013-07-24 20:00:55 & 4.7  & 166 $\pm 13$\\
00032619018 & 2013-07-31 06:55:12 & 2013-07-31 18:35:55 & 5.0  &241 $\pm 16$ \\
00032619019 & 2013-08-07 00:48:43 & 2013-08-07 22:04:55 & 4.3  & 192 $\pm 14$\\
00032619020 & 2013-08-14 15:33:40 & 2013-08-14 19:12:55 & 4.6  &207 $\pm 15$\\
\hline
\end{tabular}
\end{minipage}}
\raggedright{The source coordinates (J2000) are $\rm R.A. = 00^h15^m28\fs90$ and $\rm Decl.=-39^\circ13'19\farcs0$ \citep{stobbart06}; the Galactic column density along its direction is \nh$_{Gal}=$ 1.37$\times 10^{20}$ cm$^{-2}$ \citep{kalberla05}.}
\end{table}

\section{Data reduction}
\label{sect2}

\subsection{\cxo\ observations}

We analysed the only two archival \textit{Chandra} observations of \src\ (see Table\,\ref{logs}), taken in 2001 September and 2004 June with the Advanced CCD Imaging Spectrometer Imaging array (ACIS-I; \citealt{garmire03}). The CCDs were operated in Timed Exposure mode and in full frame (readout time: 3.2~s), and the Faint telemetry mode was used. The data were reprocessed with the \cxo\ Interactive Analysis of Observations software  package (\textsc{ciao}, version 4.6; \citealt{fruscione06}) and the calibration files in the \textsc{caldb} release 4.5.9.

Despite the source \src\ was not at the aimpoint (4' and 7' off-axis in the first and second observation, respectively), in both observations the count rate was high enough to make pile-up effects in the ACIS detector not negligible. A `pile-up map' created with the \textsc{ciao} tool \textsc{pileup$\_$map} confirmed the presence of pile-up.
Owing to the sharp \cxo\ point-spread function (PSF), correcting for pile-up discarding the data in the core of the PSF would result in a strong loss of counts. Therefore we used instead the {\sc pile-up} model by \citet{davis01} (see section~\ref{spectral_results}). The \textit{Chandra} spectra should be treated carefully: since the pile-up strongly depends on count rate, the pile-up fraction was $\sim20\%$ in the first observation and $\sim40\%$ in the second.  
We considered the pile-up level of the latter observation too high. 
The spectral modifications may be large enough that also the convolution model by \citet{davis01} may not correctly describe them, making the spectral parameters not reliable. Therefore we splitted the second observation in two parts with the same duration and we spectrally re-analyzed the second part, when the source count rate was lower and pile-up was mitigated.  
The count rate was 0.34 cts s$^{-1}$ and the pile-up grade migration ($\alpha$) decreased to $\sim0.1-0.2$, although these values are poorly constrained (see Table~\ref{tabel_best_fit}).
\begin{figure*}
\begin{center}
\includegraphics[height=7.5cm,width=8.5cm]{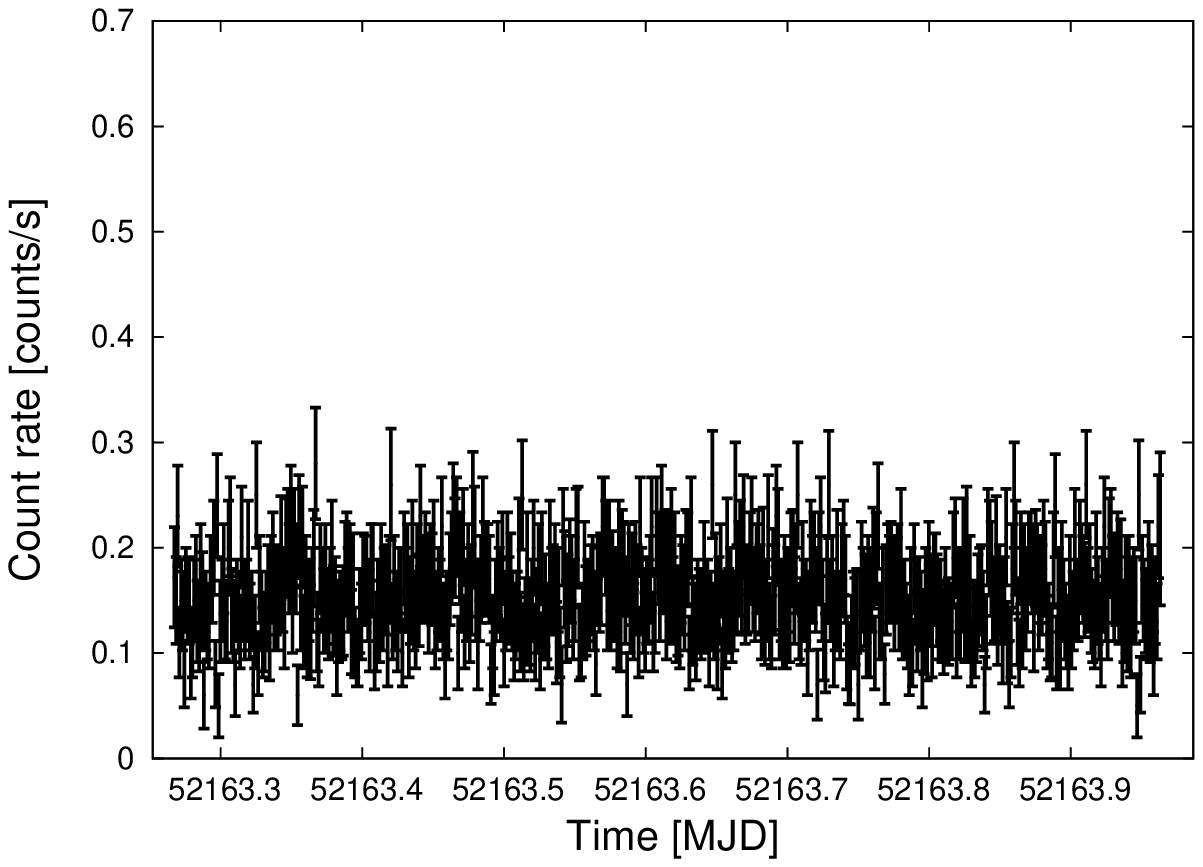}
\includegraphics[height=7.5cm,width=8.5cm]{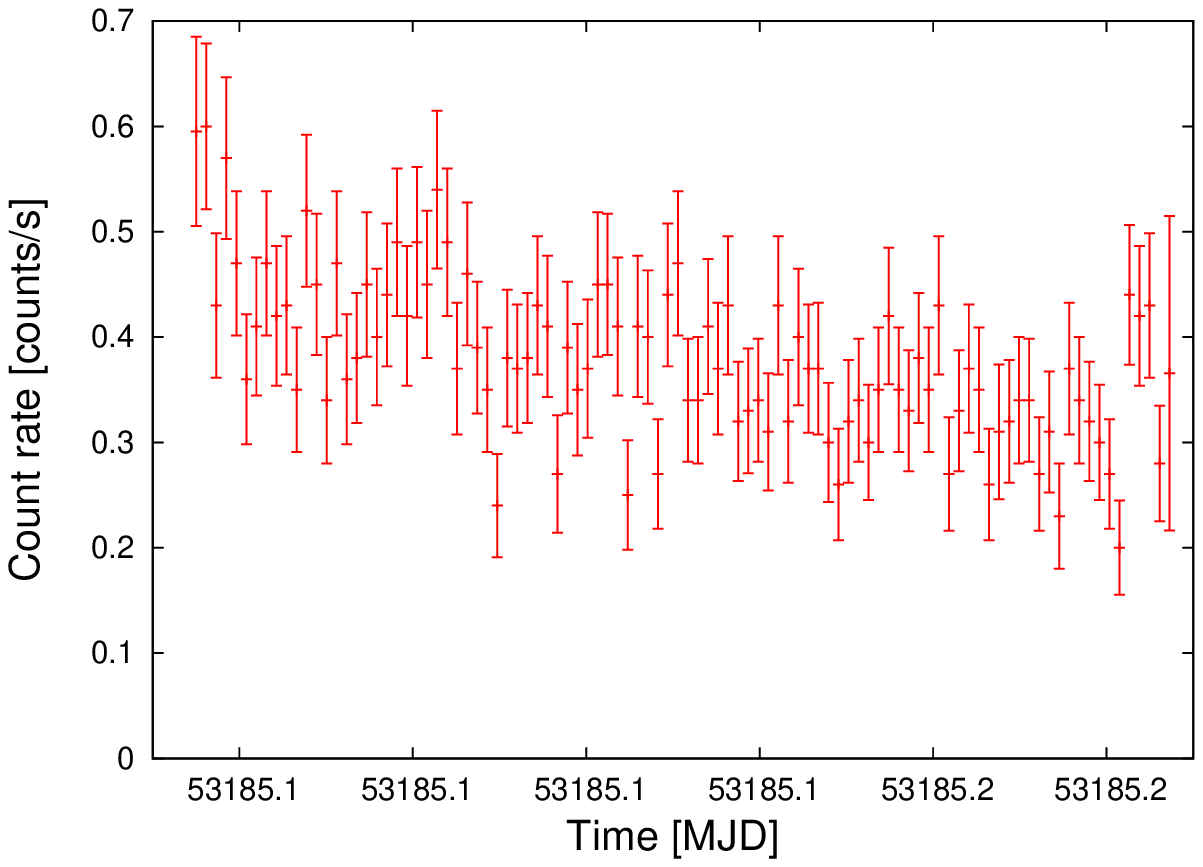}
\caption{Light curve of the two Chandra observations (Obs.ID. 2255 and 4744 at the left and right, respectively) in the 0.3--10 keV energy band, sampled with a bin size of 100s.}
\label{hardness}
\end{center}
\end{figure*}

In observation 2255, the source counts were extracted from an elliptical region with semi-axes of 5$''$ and 3$''$, while the background was estimated from an annulus with radii 8$''$ and 15$''$. In observation 4744, the semi-axes of the elliptical source region were  6$''$ and 9$''$, and the radii of the background annulus 15$''$ and 25$''$. Background regions were selected close to the source and we verified that the selected CCD regions were free of emitting X-ray sources. 

\begin{figure}
\begin{center}
\includegraphics[height=6.8cm,width=8.9cm]{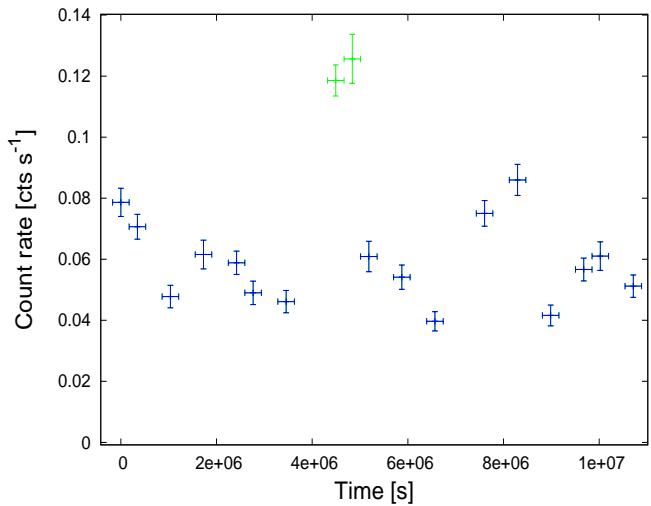}
\caption{\textit{Swift} light curve from the whole data set, with bin size of 4 days. We show the observations stacked in the \textit{Swift} ``group 1'' and ``group 2'' spectra (see text) with \textit{blue} and \textit{green} colours, respectively.}
\label{tot_lc_swift}
\end{center}
\end{figure}

\subsection{\swift\ observations}
We monitored \src\ between 2013 April and August using the X-Ray Telescope (XRT; \citealt{burrows05short}) on-board \swift. We collected approximately one 5-ks long observation per week in `Photon Counting' mode (PC, the XRT full imaging mode, which has a readout time of 2.507~s; see \citealt{hill04short}).

The XRT data were processed with \textsc{xrtpipeline} (in the \textsc{heasoft} software package version 6.14), filtered and screened with standard criteria. The source counts were extracted within a 20-pixel radius ($\sim47''$; one XRT pixel corresponds to $\sim$2.36 arcsec\footnote{http://swift.gsfc.nasa.gov/analysis/xrt\_swguide\_v1\_2.pdf}), while the background was estimated from an annulus with internal (external) radius of 60 (100) pixels, i.e. $\sim141''(236'')$. We verified that the background regions were in a CCD region free of other X-ray sources.
For the spectroscopy, we used the standard \textsc{caldb} spectral redistribution matrices, while the ancillary response files were generated with \textsc{xrtmkarf}, which accounts for different extraction regions, vignetting and point spread function corrections. 

The lightcurves were corrected using the \textsc{xrtlccorr}  tool, which takes into account Point Spread Function (PSF) corrections, vignetting and photon losses due to bad pixels/columns inside the extraction region. The latter corrections are based on a raw instrument map, constituted by a number of extensions corresponding to different time intervals, and its spatial dimension is arranged to cover the extraction region.

The statistics of each individual \emph{Swift} observation is poor and did not allow us to well constrain the spectral properties of NGC 55 ULX1 adopting the models described in the next sections. However, the hardness ratio (1.5-10 keV/0.3-1.5 keV ) did not provide indications of statistically significant spectral variability between observations. Therefore,  in order to improve the Signal-to-Noise ratio, we stacked all Swift observations according to their flux. We obtain two spectra (labeled as Swift ``group 1'' and ``group 2'') comprising 18 low flux and 2 high flux (Obs. Id. 9009 and 9010) observations, respectively (see the lightcurve in Figure 1).

\begin{figure*}
\begin{center}
\includegraphics[height=6.0cm,width=5.8cm]{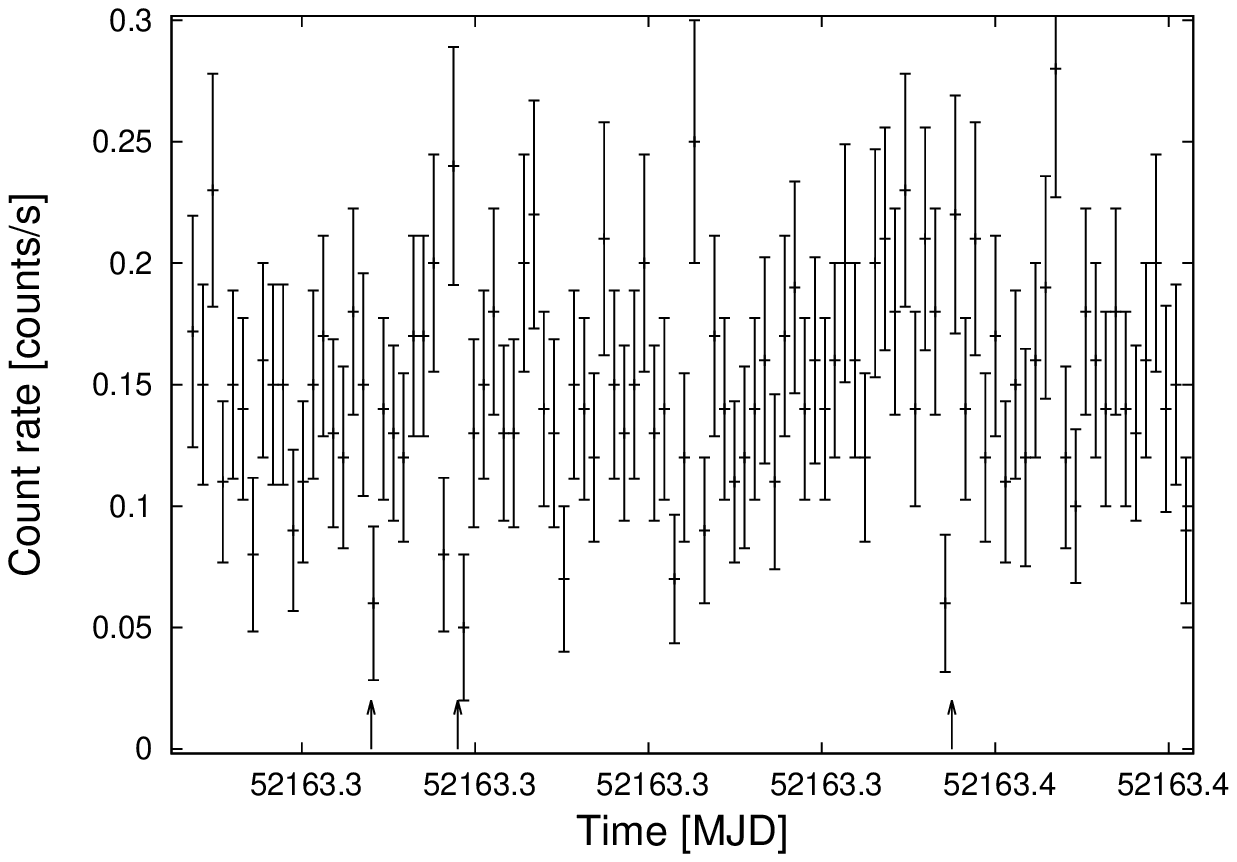}
\includegraphics[height=6.0cm,width=5.8cm]{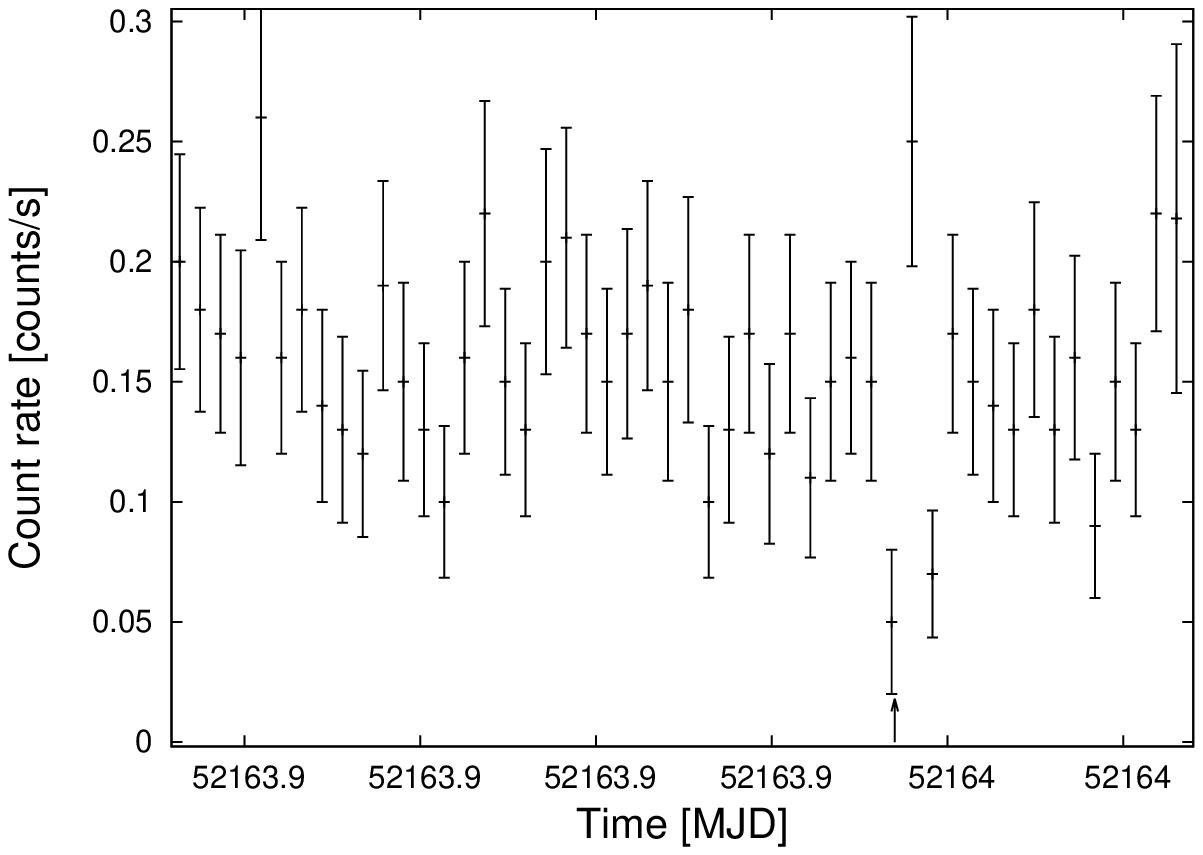}
\includegraphics[height=6.0cm,width=5.8cm]{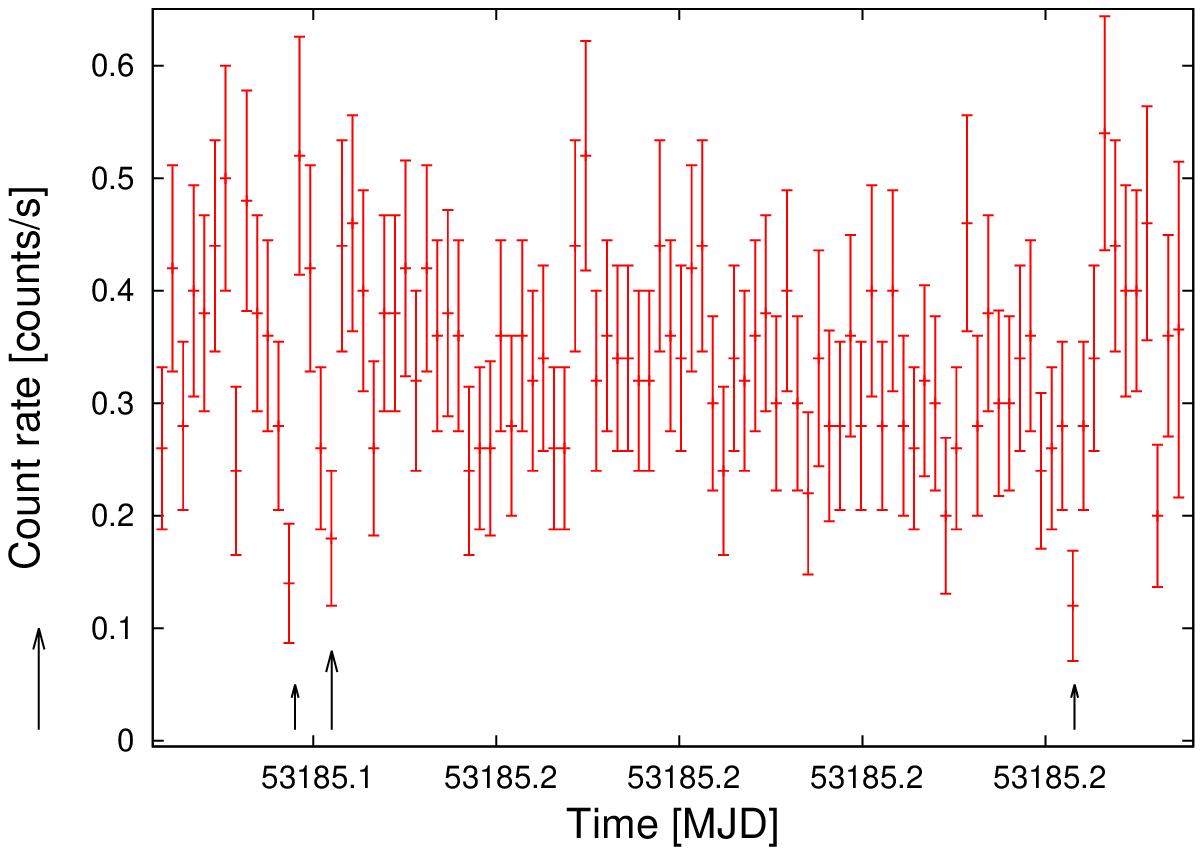}
\caption{A zoom of the light curve, in the 0.3--10 keV energy band, sampled with a bin size of 100 seconds in the first (\textit{left, center}) and 50 seconds in the second (\textit{right}) \textit{Chandra} observation. There may be marginal hints of dips at the times marked with black arrows. 
}
\label{dip_chandra}
\end{center}
\end{figure*}

\begin{figure*}
\begin{center}
\subfigure{\includegraphics[height=6.0cm,width=5.8cm]{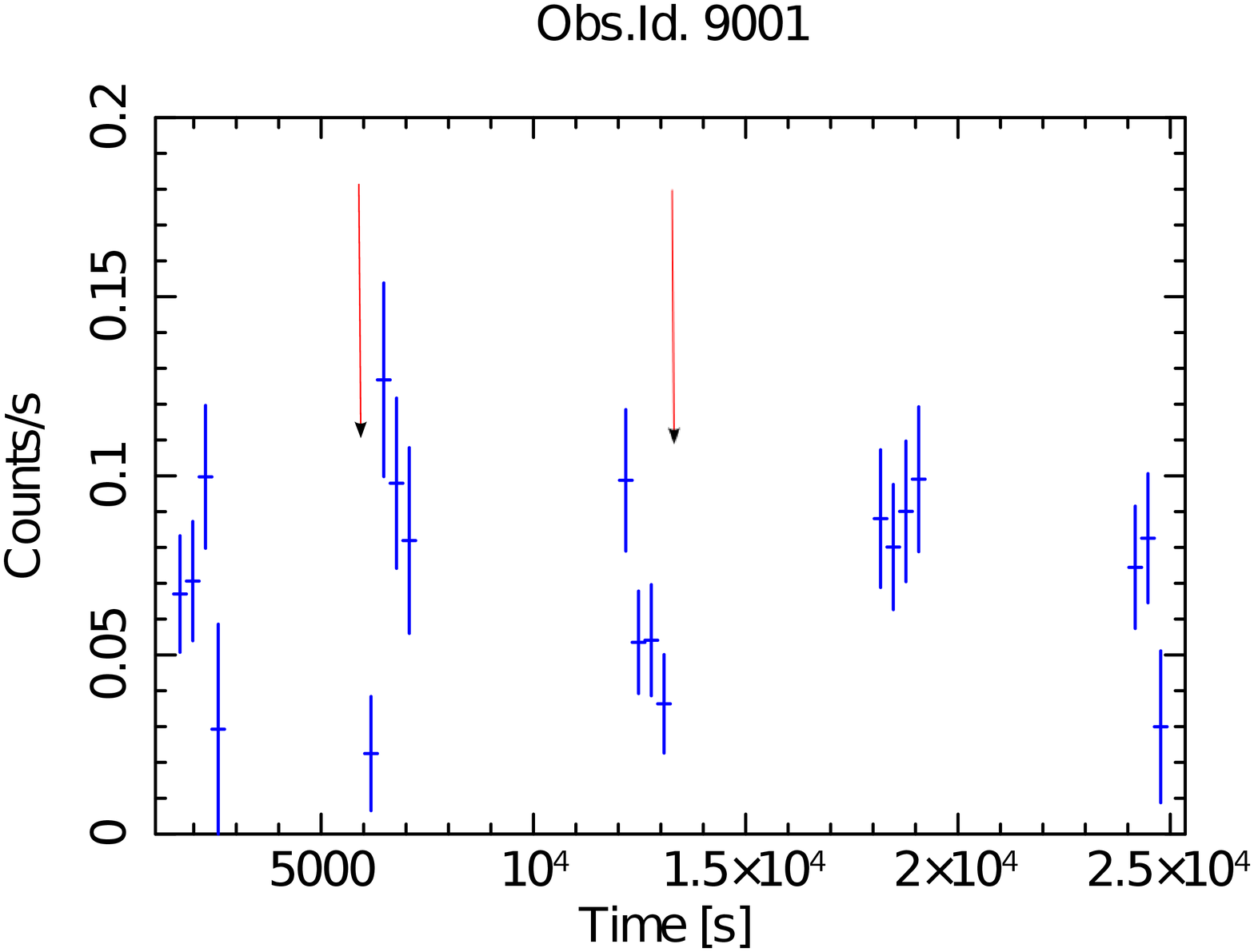}}
\subfigure{\includegraphics[height=6.0cm,width=5.8cm]{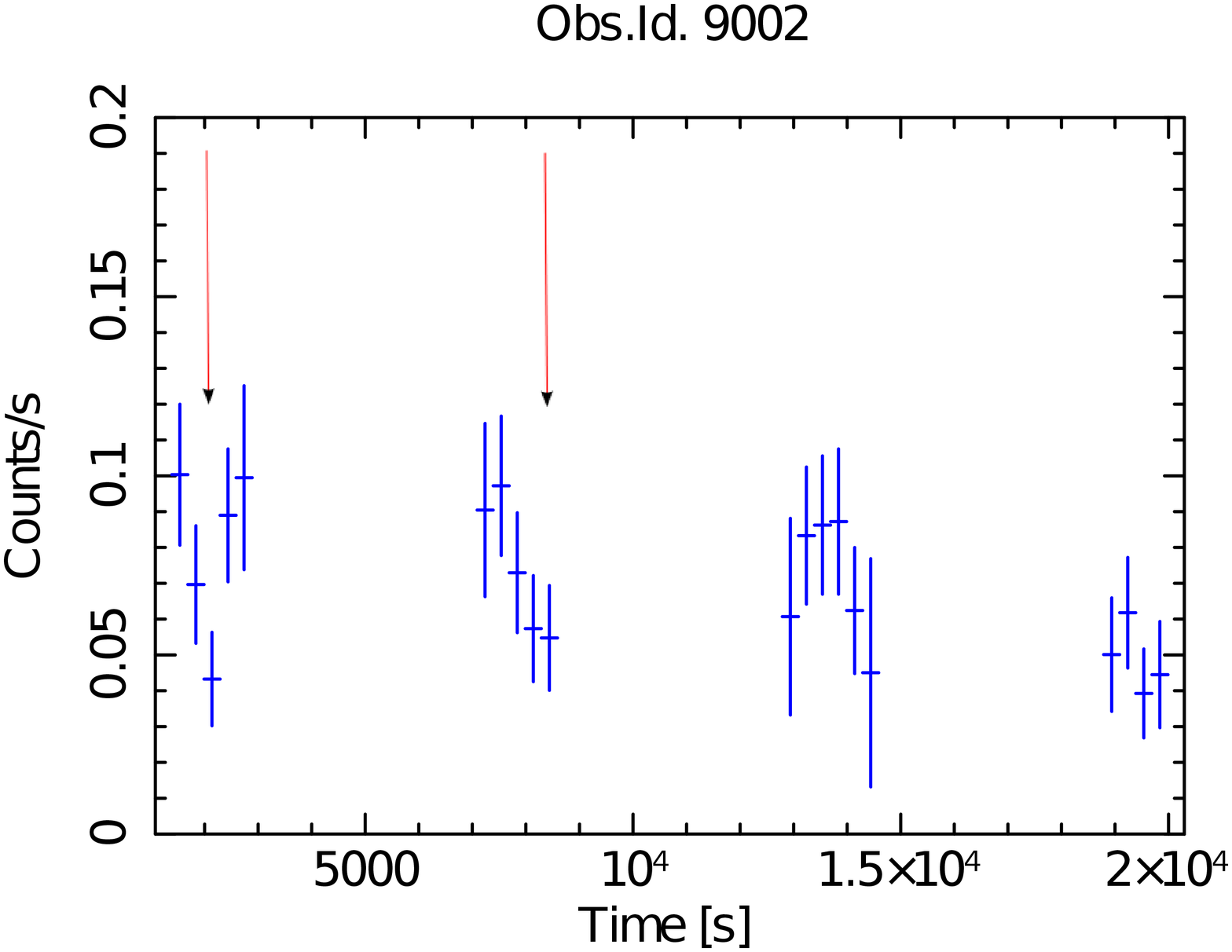}}
\subfigure{\includegraphics[height=6.0cm,width=5.8cm]{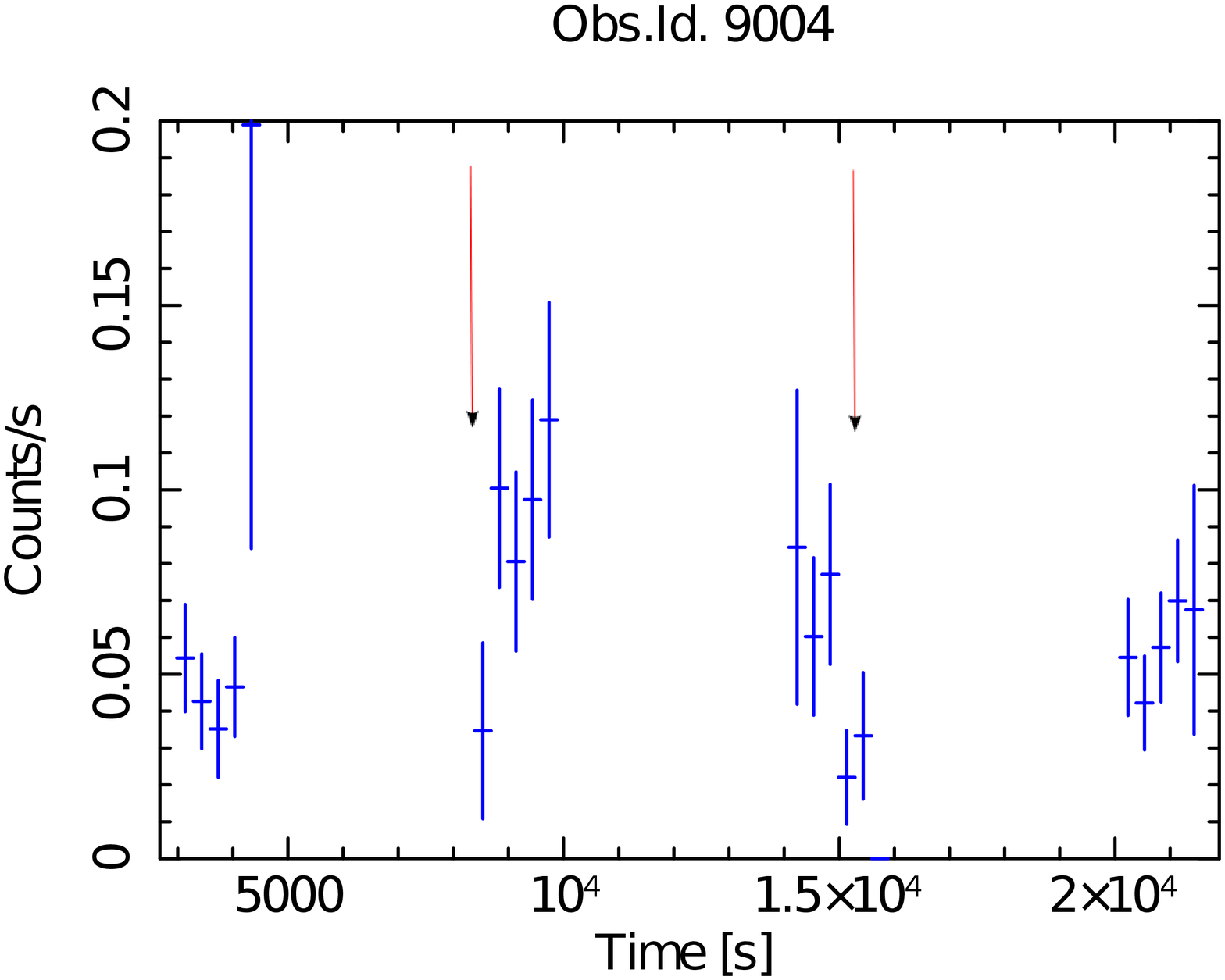}}
\caption{An example of three \emph{Swift} observations in which marginal hints of dip episodes are seen (indicated by red arrows). The bin time of the lightcurves (in 0.3--10 keV energy band) is 300 seconds.}
\label{dip_plot}
\end{center}
\end{figure*}

\section{Data analysis}
\label{spectral_results}

\subsection{Timing analysis}
{A peculiar property of NGC 55 ULX1 is the existence of narrow dips in its \emph{XMM-Newton} lightcurves. { \citet{stobbart04} showed that, on average, the dips last about 100-300s (although a few dips appeared to be longer), with a depth at the level of $\sim80-90\%$ of the persistent flux ($\sim2.5\times10^{-12}$ erg cm$^{-2}$ s$^{-1}$) and more intense in the 2.0--4.5 keV energy band. We note that the average observed flux of NGC 55 ULX1 in the \emph{Chandra} and \emph{Swift} observations is $\sim1.5\times10^{-12}$ erg cm$^{-2}$ s$^{-1}$, therefore lower than in the \emph{XMM-Newton} data. If dips are characteristic of a specific high flux state of the source, we might not be able to detect any dip in the \emph{Chandra} and \emph{Swift} datasets. However, we can suppose that they are a random effect and present at different flux levels.
In the \textit{XMM-Newton} observations, we counted around 30 dips in a total exposure time of 51.9 ks. As they represent a small fraction of the total \textit{XMM-Newton} exposure time, we calculated the expected number of dips detections $N_{dips}$ in the \textit{Chandra} and \textit{Swift} observations according to the relation: $N_{dips}=(N_{\text{XMM}}/ t_{\text{XMM}}) \times t_{exp}$, where $N_{\text{XMM}}$ is the total number of dips in the \textit{XMM-Newton} observations, $t_{\text{XMM}}$ is the total \textit{XMM-Newton} exposure time and $t_{exp}$ is the \textit{Chandra} or \textit{Swift} exposure time. We therefore expected to detect $\sim 34$ dips in the first \textit{Chandra} observation (59.4 ks), $\sim 5$ dips in the second \textit{Chandra} observation (9.6 ks), 2-3 dips in any single \textit{Swift} snapshot ($\sim4.5$ ks) or $\sim 55$ dips during the total \textit{Swift} exposure time (95 ks).

The \textit{Chandra} observations showed an increase in the count rate of about a factor of 2-3 between the two epochs (Fig.~\ref{hardness}), and an average count rate in the 0.3--10 keV energy band of $\sim0.15$ cts$^{-1}$ and $\sim 0.4$ cts s$^{-1}$ for the first and second observation, respectively. Assuming to observe occultations similar to those of the \textit{XMM-Newton} observations (depth at the level of $\sim80-90\%$ of the persistent flux), we should detect dips at $\sim0.03$ and $\sim0.1$ cts s$^{-1}$ in the first and second \textit{Chandra} observation, respectively. We sampled the 0.3--10 keV lightcurve of the first \textit{Chandra} observation with 100s time bins as it assured the minimum binning to obtain Gaussian statistics in each time bin. We found only marginal indications of a limited number of dips, although the uncertainties due to the poor quality of the data are large ($\sim$$40\%$ at the lowest rates; see some examples in Figure~\ref{dip_chandra}-- \textit{left, center}). The lightcurve of the second \textit{Chandra} observation was instead sampled with 50s, as it assured the minimum binning to obtain Gaussian statistics in each time bin. However, also in this case, we found only marginal evidence of dips ($\sim$$60\%$ error bars; see Figure~\ref{dip_chandra}-\textit{right}).} 

The \emph{Swift} observations generally showed variability of less than a factor of 2 (except for two high flux observations, { that reached $\sim2.4\times10^{-12}$ erg cm$^{-2}$ s$^{-1}$;} Fig.~\ref{tot_lc_swift}). { Assuming a persistent mean rate between $\sim$0.06 and $\sim$0.1 cts s$^{-1}$, we expected the depth of the dips to be at $\sim 0.01-0.02$ cts s$^{-1}$, with a corresponding error bar of $\sim$60-80$\%$. The relatively low count rate detected by \textit{Swift} did not allow us to investigate the existence of dips on timescales lower than 300s, as it represents the best compromise to reach gaussian statistics in almost each time bin and have small enough error bars. However, this implies that only the longest dips could be detected in these datasets.
None of such deep dips at $\sim 0.02-0.01$ cts s$^{-1}$ was reliably detected, although we found possible hints in the lightcurve of a few observations where short drops in the count rate were observed (some examples are given in Fig.~\ref{dip_plot})}. 

Dipping episodes in Galactic accreting sources are usually characterised by a spectral variability which can be clearly revealed by inspection of the hardness ratios. However, the large error bars of the hardness ratios in our \textit{Swift} and \textit{Chandra} observations do not allow us to clearly detect them and further investigate their properties. { Finally, we checked also the existence of dips in the 2.0--4.5 keV energy band (the same band in which \citealt{stobbart04} found the strongest evidence) in both the \textit{Chandra} and \textit{Swift} observations, but the statistics was too poor.}}

\begin{table*}
\footnotesize
\begin{center}
   \caption{Best fitting spectral parameters of NGC 55 ULX-1 obtained with an absorbed \textsc{bbody+diskbb} model. The error bars refer to the 90$\%$ confidence level.}
\scalebox{0.8}{\begin{minipage}{22cm}
   \label{tabel_best_fit}
   \begin{tabular}{l c c c c c c c c c}
\hline
&N$_\mathrm{H}$$^a$ & kT$_\mathrm{bb}$$^b$ & kT$_\mathrm{disc}$$^c$ & $f_\mathrm{obs}$ [0.3--10 keV]$^d$ & $L_\mathrm{X}$ [0.3--10 keV] $^e$& $L_\mathrm{diskbb}$[0.3--10 keV] $^f$&$\chi^2/$dof &$\alpha$$^g$& $f^h$\\
& $10^{22}$ atoms cm$^{-2}$ & keV & keV & $10^{-12}$ erg cm$^{-2}$ s$^{-1}$ & $10^{39}$ erg s$^{-1}$ & $10^{39}$ erg s$^{-1}$ & & &$\%$\\
\hline
\textit{Chandra} (Obs.ID. 02255) & $0.24_{-0.08}^{+0.09}$ & $0.18_{-0.02}^{+0.03}$ & $0.64_{-0.05}^{+0.05}$& 1.10$_{-0.10}^{+0.10}$ &$1.3_{-0.20}^{+0.10}$& $0.6_{-0.1}^{+0.1}$& 141.44/149 & 0.20$_{-0.10}^{+0.10}$ & 0.85\\
\textit{Chandra} (Obs.ID. 04744)$^*$ & $0.30_{-0.30}^{+0.40}$& $0.15_{-0.03}^{+0.09}$ & $0.61_{-0.10}^{+0.08}$& 1.70$_{-0.20}^{+0.15}$  &$14.0_{-2.0}^{+5.0}$& $7.0_{-1.0}^{+3.0}$&  46.19/40 & $0.13_{-0.13}^{+0.23}$&0.85\\

\textit{Swift}, ``Group 1''  & $0.60_{-0.10}^{+0.10}$& $0.13_{-0.02}^{+0.02}$ & $0.56_{-0.04}^{+0.04}$&  1.11$_{-0.03}^{+0.03}$ &$3.8_{-0.40}^{+1.30}$& $3.0_{-1.0}^{+2.0}$ & 148.90/124 &$-$ &$-$\\
\multirow{2}{*}{\textit{Swift}, ``Group 2'' } & $0.13_{-0.06}^{+0.07}$& $-$ & $0.60_{-0.06}^{+0.06}$& 1.50$_{-0.20}^{+0.20}$  &$1.1_{-0.10}^{+0.20}$& $1.1_{-0.1}^{+0.2}$ & 39.45/34 & $-$&$-$\\
 & $0.49_{-0.04}^{+0.05}$& $0.13(f)^{**}$ & $0.59_{-0.06}^{+0.06}$& 2.40$_{-0.20}^{+0.10}$  &$2.8_{-0.20}^{+0.20}$& $1.3_{-0.1}^{+0.2}$ & 41.42/34 & $-$&$-$\\
\hline
\end{tabular} 
\end{minipage}}
\end{center}
\begin{quotation}\footnotesize
$^a$ Intrinsic column density of the neutral absorber; $^b$ Temperature of the blackbody component; $^c$ Inner disc temperature; $^d$ Observed flux in the 0.3--10 keV energy range; $^e$ Unabsorbed luminosity in the 0.3--10 keV energy range, assuming a distance of 1.78 Mpc; $^f$ Unabsorbed disc luminosity in the 0.3--10 keV energy range, assuming a distance of 1.78 Mpc; $^{g,h}$ Parameters of the \textsc{jdpileup} pile-up model as implemented in \textsc{XSPEC}: $\alpha$ is the grade-migration parameter (the probability that $n$ events will be piled together but will still be retained after data filtering is $\alpha^{n-1}$); $f$ is the fraction of the PSF treated for pile-up and is required to be in the range 85--100\%; \\
$^*$ This observation has been split in two parts because of high pile-up; here we consider only the second part, where pile-up is lower.
$^{**}$ Alternative fit where the parameters of the {\sc bbody} component are fixed to those found in the ``group 1'' \textit{Swift} spectrum. 
\\ 
\end{quotation}
\end{table*}

\subsection{Spectral analysis}
We analysed all the \textit{Chandra} and \emph{Swift} spectra obtained as described in Section~\ref{sect2} using {\sc {\sc XSPEC}} v. 12.6.0 \citep{arnaud96} and grouping them to obtain at least 20 counts per bin. Two absorption components (\textsc{tbabs} in {\sc XSPEC}) were considered in all spectral fits: the first was introduced to model the Galactic absorption along the direction of the source and fixed at $1.37\times10^{20}$ cm$^{-2}$ \citep{kalberla05}. The second was allowed to vary and describes the absorption in the local environment of the source. The luminosities were estimated from the flux calculated from the \textsc{cflux} convolution model in {\sc XSPEC}, assuming a distance of 1.78 Mpc \citep{karachentsev03}.

From the analysis of the \emph{XMM-Newton} data (\citealt{gladstone09,sutton13}), NGC 55 ULX1 showed a complex continuum that can be described by a two-components model, consisting of a multicolour blackbody disc and a Comptonization component with a roll-over at high energy. 
However, this model was difficult to apply to the \textit{Chandra} and the stacked \textit{Swift} spectra because the data quality was poorer and did not allow us to constrain the spectral parameters.
We then fitted the soft component with a single blackbody model ({\sc bbody} in {\sc XSPEC}), neglecting that the emission of the outer disc/wind may come from layers with different ionisation and temperature. For similar reasons, the broad high energy spectrum, often described in terms of a Comptonization component (e.g \citealt{stobbart06,gladstone09,sutton13,pintore14}), was approximated with a multicolour blackbody disc ({\sc diskbb} in {\sc XSPEC}). Thus, for both the \textit{Chandra} and stacked \emph{Swift} spectra we adopted, as common reference model, a {\sc diskbb} \citep{mitsuda84} component plus an additional {\sc bbody} component (when needed).
\newline
\newline
\subsection{{\emph Swift} spectra} 
\begin{figure*}
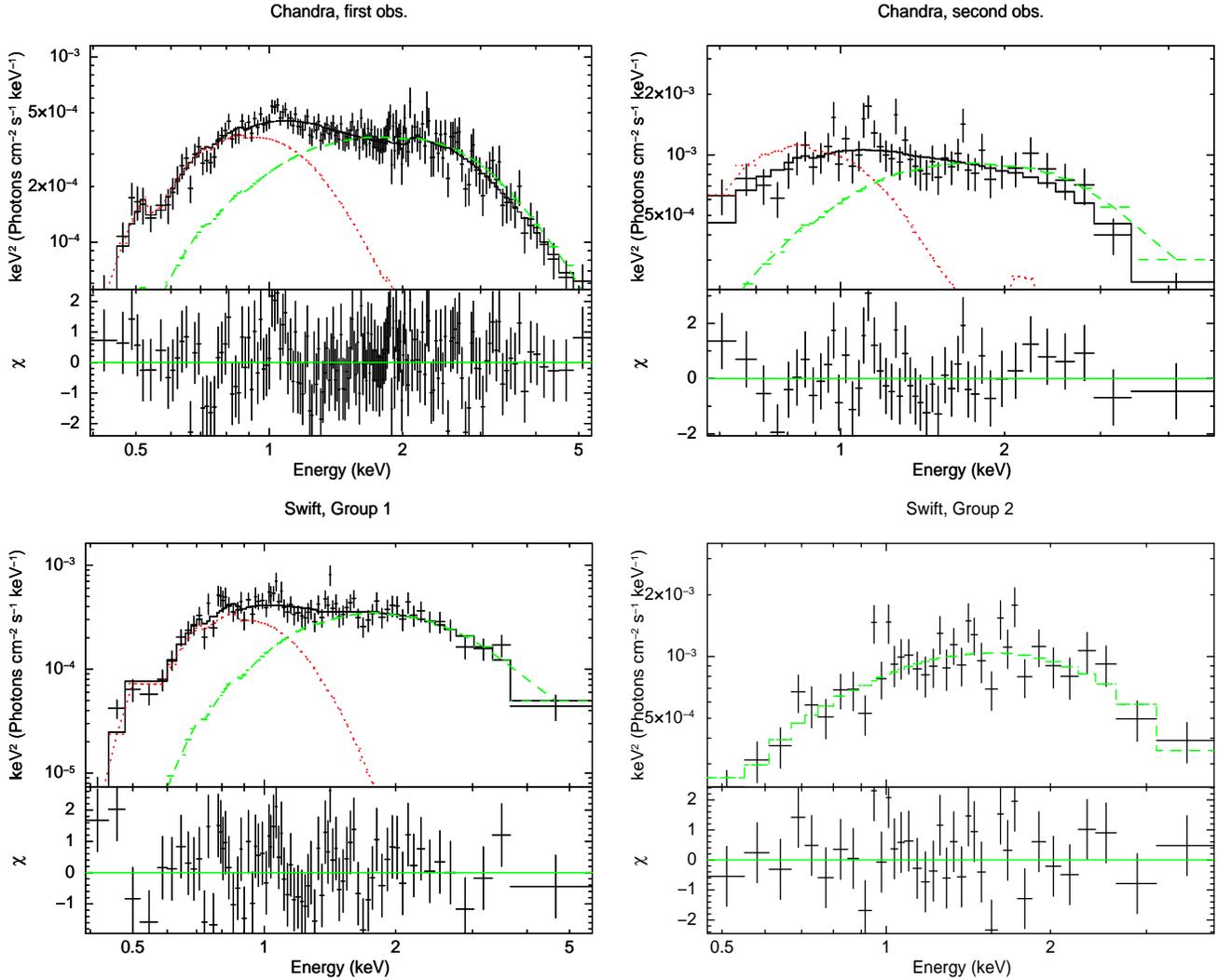

\begin{center}
\subfigure{\includegraphics[height=8.8cm,width=6.9cm,angle=270]{chandra-first_obs.eps}}
\subfigure{\includegraphics[height=8.8cm,width=6.9cm,angle=270]{chandra-second_obs_second_part.eps}}
\subfigure{\includegraphics[height=8.8cm,width=6.9cm,angle=270]{spettro_stacked_low_flux.eps}}
\subfigure{\includegraphics[height=8.8cm,width=6.9cm,angle=270]{spettro_stacked_high_flux.eps}}
\caption{\textit{Top}: unfolded spectra ($E^2f(E)$) of the first (\textit{top-left}) and second (\textit{top-right}) Chandra observations; \textit{bottom}: unfolded spectra ($E^2f(E)$) of the stacked ``group 1'' (\textit{left}) and ``group 2'' (\textit{right}) \textit{Swift} observations. The continuum is described by an absorbed blackbody ({\sc bbody}, dotted, red line) plus a multicolour disc ({\sc diskbb}, dashed, green line) or a simple {\sc diskbb} (see text). }
\label{spectra}
\end{center}
\end{figure*}
The ``Group 2'' spectrum does not need an additional {\sc bbody} component, as a single disc component, with a temperature of 0.6 keV provides a good fit to the data (see Table~\ref{tabel_best_fit}). We note also that a {\sc diskbb} model is marginally favoured in comparison with a single \textsc{powerlaw} component ($\chi^2$/dof=44.39/34). Because of the disc-like spectral shape, we tried to fit the ``Group 2'' spectrum also with a modified disc model ({\sc diskpbb}, \citealt{mineshige94}), in which the radial dependence of the temperature goes as T$_\mathrm{disc}\propto r^{-p}$, and \textit{p} is 0.75 for a standard disc and 0.5 for a slim disc. This model  provided a good fit but its parameters turned out to be totally unconstrained. 

The best fit of the ``Group 1'' spectrum was instead obtained with a two-components model, a blackbody with a temperature of 0.13 keV and a high energy {\sc diskbb} component with a temperature of 0.56 keV (see Table~\ref{tabel_best_fit}). As the { mean flux estimated from} the ``Group 2'' spectrum was about a factor of 4 lower than that of the ``Group 1'' \emph{Swift} spectrum, we tested the possibility that the soft blackbody component might not be detected in the "Group 2" spectrum because of the poor counting statistics. We then added to the {\sc diskbb} component a {\sc bbody} model with fixed spectral parameters, equal to those found in the ``group 1'' spectrum (see below) and letting the column density, the {\sc diskbb} temperature and normalization free to vary. We found that the additional component was acceptable and led only to a change in column density which grew from $0.13$ atoms cm$^{-2}$ to $0.49$ atoms cm$^{-2}$ (Table~\ref{tabel_best_fit}).

Comparing the spectral parameters of the two \textit{Swift} spectra, we find that, while the total absorbed 0.3--10 keV luminosity increased (from $5-9\times10^{38}$ erg s$^{-1}$ to $1\times10^{39}$ erg s$^{-1}$) with the mean count rate of the stacked spectra, the unabsorbed luminosity did the opposite. The intrinsic source luminosity (i.e. unabsorbed) was about a factor of 4 higher in the ``group 1'' spectrum than in the ``group 2'' spectrum ($\sim4\times10^{39}$ erg s$^{-1}$ vs $\sim1-3\times10^{39}$ erg s$^{-1}$). This may be explained in terms of the higher column density (\nh) of the ``group 1'' spectrum ($\sim 0.6\times 10^{22}$ cm$^{-2}$), about a factor of 6 higher than in the ``group 2'' spectrum ($\sim0.1\times10^{22}$ atoms cm$^{-2}$). However, comparing with the alternative two-components fit of the ``group 2'' spectrum, the luminosity of the ``Group 1'' spectrum is only a factor 1.5 higher and the column density only 20\% larger.
\newline
\newline
\subsection{{\emph Chandra} spectra} 
\begin{figure}
\begin{center}
\includegraphics[height=6.3cm,width=8.8cm]{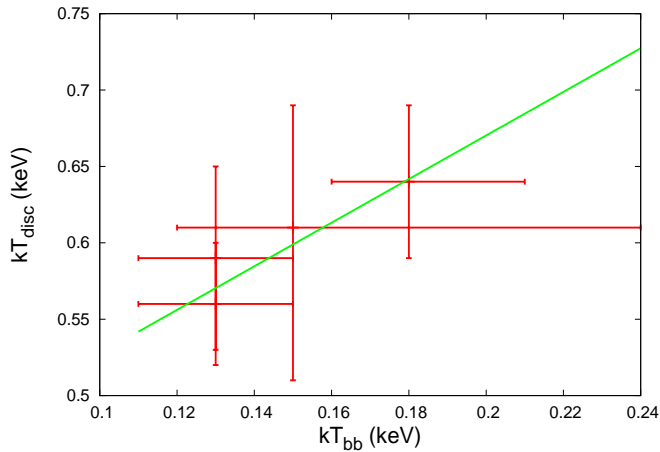}
\caption{Tentative linear correlation between the blackbody and disc temperature (taking into account the alternative fit of the ``group 2'' spectrum).}
\label{ktdsicvsktbb}
\end{center}
\end{figure}

The parameters of the {\sc diskbb} and {\sc bbody} components of the first observation are kT$_\mathrm{bb}\sim0.18$ keV and kT$_\mathrm{disc}\sim0.64$ keV, slightly higher than those of the \emph{Swift} spectra (but consistent with them within the errors). The spectral parameters of the second observation are less constrained. The column density is consistent with 0 within the errors, while the temperatures of the {\sc bbody} and {\sc diskbb} components ($\sim0.15$ and $\sim0.6$ keV) are comparable to those of the first observation. We note that the intrinsic (i.e. unabsorbed) luminosity of the second {\emph Chandra} observation is clearly the highest of the sample, an order of magnitude higher than the intrinsic luminosity of the ``group 2'' \textit{Swift} spectrum.
\newline
\newline
\noindent There is some hint of spectral variability between the \emph{Swift} and \textit{Chandra} observations, although most of it is accounted for by variations in column density (spanning the range between $\sim$1.5$\times10^{21}$ atoms cm$^{-2}$ and 6$\times10^{21}$ atoms cm$^{-2}$) and normalization of the spectral components.
Using the alternative two-components fit of the ``group 2'' spectrum, we find marginal evidence of a linear correlation between the temperatures of the blackbody and disc components (kT$_{disc}=(1.4\pm0.4)$ kT$_{bb}+(0.38\pm0.06)$; see Fig.~\ref{ktdsicvsktbb}), with a regression coefficient of 0.92 and a Spearman coefficient of 0.8. This may suggest that the evolution of the two components is coupled, although it is not clear how it relates to changes in the mass accretion rate.

Finally, in both the {\it Swift} and {\it Chandra} spectra, there is some indication (not statistically significant, though) of an excess at $\sim 1$ keV in the residuals, that may be associated to an emission feature. Other ULXs showed a similar puzzling feature in their spectra, that has been interpreted as emission from a plasma close to the source (i.e. \citealt{strohmayer07,middleton11,caballero13,sutton13,pintore14}) or mismodeling of absorption features (\citealt{middleton14}a).

\section{Discussion and conclusions}
\label{discussion}

We analysed two \textit{Chandra} ACIS-I observations and 20 \emph{Swift} observations, taken in PC mode, of the Ultraluminous X-ray source NGC 55 ULX1. We investigated the spectral variability of the source, stacking the \textit{Swift} observations in two spectra according to the observed count rate, with the aim to improve the signal-to-noise ratio. For both the stacked \textit{Swift} spectra and the \textit{Chandra} spectra we adopted a model consisting of a blackbody (for the soft component) and a multicolour accretion disc (for the hard component). The \textit{Chandra} and \textit{Swift} spectra showed spectral variability similar to that observed in other ULXs (e.g. \citealt{kajava09,feng09,sutton13,pintore14}). We can account for this spectral variability mostly in terms of changes in both normalization and column density of the two model components.

Despite detailed timing information are missing, as many other ULXs, \src\ escapes an interpretation in terms of transitions between canonical sub-Eddington states as those observed in accreting Galactic BH X-ray binaries (e.g. \citealt{mcclintock06}). Although different scenarios have been proposed (e.g. \citealt{miller14}), many of the observed properties can be qualitatively accounted for by the expected behaviour of an accretion disc accreting above Eddington \citep[e.g.][]{gladstone09,zampieri09,middleton12,sutton13,pintore14}. In such conditions, advection of energy takes place in the accretion disc, the structure of which is modified with respect to that of a standard disc (\citealt{abramowicz88,watarai01,ohsuga11}). In addition, part of the accretion energy is removed in forms of radiative outflows that originate in the inner regions of the disc \citep[e.g.][]{poutanen07,ohsuga09}. Their opening angle (measured from the rotation axis of the disc) decreases when the accretion rate increases \citep[e.g.][]{takeuchi14}. The density of these outflows is sufficiently high that they are optically thick and produce thermal emission. The characteristic temperature of the photosphere in such winds is consistent with that of the soft component often detected in ULX spectra. If the angle between our line of sight and the rotation axis of the accretion disc is sufficiently large, short term variability \citep{heil09} can be induced by optically thick clouds of matter that intersect the edge of the winds \citep{middleton11,sutton13}. In case the wind is sufficiently dense and extended, its outer neutral phase can also partially obscure the high energy emission from the inner disc and its own soft emission, leading to an increase in the local absorption and column density \citep[e.g.][]{pintore14}. Narrow absorption lines from this warm phase of the wind are still not yet convincingly detected (\citealt{walton12}), although broad, {blue-shifted X-ray emission features of ionised species of abundant elements may be expected (\citealt{middleton14})a}.
In the following, we then interpret the observed spectral evolution of \src\ as induced by variations in accretion rate, outflow rate and intrinsic absorption assuming that the source is in a super-Eddington accretion state of this type and that the inclination angle to the source is large.

We then start from the second \textit{Chandra} observation, the intrinsic luminosity of which reaches $10^{40}$ erg s$^{-1}$ (Table \ref{tabel_best_fit}), a value higher than that of any \textit{XMM-Newton} observation. Both the soft and hard emission are stronger than in any previous observation. We then suggest that, at the time of this observation, the accretion rate is the highest, and that the outflow is extremely powerful with a very extended photosphere. In the first \textit{Chandra} observation the spectrum shows still evidence of a powerful soft component, but with a total intrinsic X-ray luminosity of $\sim 1.3\times 10^{39}$ erg s$^{-1}$. The accretion rate is then probably lower than in the second \textit{Chandra} observation. The ``group 1'' \textit{Swift} observations represent a sort of intermediate state in which the accretion rate is probably smaller than that of the second \textit{Chandra} observation, but higher than that in the first.
While the outflow was still strong, the ionising flux from the disc decreased and then the fraction of neutral wind material became higher, absorbing more effectively its own emission and that of the inner regions. We note that the column density of these observations is $\sim 0.6\times 10^{21}$ erg s$^{-1}$, the highest amongst all the spectra considered here. The absorbed luminosity (and count rate) is so low because the view of the central regions is highly suppressed by this obscuring low-ionisation component. Finally, the interpretation of the state of the source during the ``group 2'' \textit{Swift} observations depends somewhat on the spectral fit considered (see Table~\ref{tabel_best_fit}). Assuming the single component disc fit, the intrinsic low luminosity and intrinsic column density, on one side, and the lack of the soft component on the other, may suggest that the accretion rate is at the lowest level. According to numerical simulations \citep[e.g.][]{poutanen07,ohsuga09}, this would represent a state in which the wind photosphere is not extended and the wind opening angle decreases to the point that it is no longer intersecting our line of sight towards the central regions. Considering the two-components fit, the situation is less clear, but may be suggestive of a relatively weak wind with a still well developed neutral fraction. As the two ``group 2'' \textit{Swift} observations were collected within 20 days { after the first low flux interval (see Fig~\ref{tot_lc_swift})}, this second scenario may be preferable unless we allow for a total suppression of the wind in such a small timescale. 

One of the tests of the super-Eddington scenario is the detection of absorption features originating in the warm wind. The counting statistics of both the \textit{Swift} and \textit{Chandra} data did not allow us to perform a detailed analysis, but we note that there are marginally significant residuals in the spectral fits at $\sim 1$ keV in all the spectra. \citet{middleton14}a interpreted features of this type as arising in the wind. Deeper observations and higher quality spectra will help to confirm these claims.

Finally, we comment on the dips in NGC 55 ULX1. One of the aims of the \emph{Swift} monitoring was to unveil possible flux-dependent dips. In the \emph{XMM-Newton} data, dips lasting for 100-300s were observed only during the high flux states, and they were explained as caused by optically thick clouds of matter in the source environment, that from time to time encounter our line of sight \citep{stobbart04}. If the mechanism for producing them is similar to that of the Low Mass X-ray binary systems (LMXBs), then the source is seen at relatively high inclination ($>65$\textdegree, e.g. \citealt{white83}). However, although we have observed the source spanning a rather large interval in flux (a factor of 3-4 in both the \textit{Chandra} and \emph{Swift} data), no clear dip episodes were observed but only marginal drops in the flux. {Assuming occultation of 80-90$\%$ of the persistent flux, for the given exposure times of the \textit{Chandra} and \textit{Swift} observations, we would expect to detect a large number of dips. However the counting statistics of these datasets was poor, with large uncertainties ($\sim$40-60$\%$) at the flux level of the candidate dips: for the \textit{Swift} data, we could investigate only dips longer than 300s, while in the \textit{Chandra} data we could investigate shorter timescales. In some \emph{Swift} observations and in the two \textit{Chandra} observations}, we found marginal hints of narrow drops in the count rate although most of them are consistent with the mean within 3$\sigma$. An analysis of the lightcurves in the 2.0--4.5 keV energy band, where \citet{stobbart04} found the strongest evidence of dips, showed that the counting statistics was, even more, too poor to allow detection of dips.
In addition, the hardness ratios do not show any clear spectral change between dips and `persistent' episodes. 

{The lack of evidence of detectable dips in our data can therefore be attributed
either to poor counting statistics or to the random occurrence of
favourable/unfavourable obscuring conditions by blobs in the wind. The latter is
ultimately related to the unknown physical interplay between the accretion phase and
the disc/wind propagation (\citealt{middleton15}c) in NGC 55 ULX1. }Indeed, also the
spectral variability of the soft component suggests that the wind, and its
ionisation, could be highly variable on time scales from a few ks (the duration of the dips in the \textit{XMM-Newton} lightcurve) to several days (high flux \textit{Swift} observations). 

Further spectral investigations, complemented with a timing analysis, using high quality observations will allow us to further improve our understanding of \src\ and the role of outflows in ULXs.

\section*{Acknowledgments} 
We thank the anonymous referee for the useful comments and suggestions. 
F.P., P.E., L.Z., and A.W. acknowledge financial support from the INAF research grant PRIN-2011-1 (``Challenging Ultraluminous X-ray sources: chasing their black holes and formation pathways'') and the ASI/INAF contract I/004/11/0 and ASI/INAF n.
I/037/12/0. PE acknowledges a Fulbright Research Scholar grant administered by the U.S.--Italy Fulbright Commission and is grateful to the
Harvard--Smithsonian Center for Astrophysics for hosting him during his Fulbright exchange. This research is based on observations made by the \cxo\ X-ray Observatory and has made use of software provided by the \cxo\ X-ray Center (CXC, operated for NASA by SAO) in the application packages CIAO and ChIPS. This research is based on observations with the NASA/UKSA/ASI mission \textit{Swift}.
\addcontentsline{toc}{chapter}{Bibliography}
\bibliographystyle{mn2e}
\bibliography{biblio}

\label{lastpage}

\end{document}